\documentclass[letterpaper]{article} 
\usepackage{aaai2027}  
\usepackage[hyphens]{url}  
\usepackage{graphicx} 
\usepackage{natbib}  
\usepackage{caption} 
\usepackage{algorithm}
\usepackage{algorithmic}
\usepackage{enumitem}
\usepackage{newfloat}
\usepackage{listings}
\DeclareCaptionStyle{ruled}{labelfont=normalfont,labelsep=colon,strut=off} 
\floatstyle{ruled}
\newfloat{listing}{tb}{lst}{}
\floatname{listing}{Listing}
\usepackage{booktabs}
\usepackage{amsmath}
\usepackage{amssymb}
\usepackage[most]{tcolorbox}

\definecolor{main}{HTML}{4472C4}
\definecolor{sub}{HTML}{EBF4FF}
\newtcolorbox{boxA}{
  enhanced, breakable,
  boxrule = 0pt,
  colback = sub,
  borderline west = {2pt}{0pt}{main},
  borderline east = {2pt}{0pt}{main},
}

\title{Is Retrieval All You Need? Assessment and Emergence of Novelty in Protein Structure Generation}
\author{
    Tongyue Xu\textsuperscript{\rm 1,\rm 2},
    Yijie Zhang\textsuperscript{\rm 3,\rm 4},
    Mutian He\textsuperscript{\rm 5},\\
    Lingdong Shen\textsuperscript{\rm 6},
    Zhihong Liu\textsuperscript{\rm 2},
    Tianlei Ying\textsuperscript{\rm 2}\textsuperscript{*},
    Cheng Tan\textsuperscript{\rm 7}\textsuperscript{*}
}
\affiliations{
    \textsuperscript{\rm 1}Tongji University,
    \textsuperscript{\rm 2}Shanghai Innovation Institute,
    \textsuperscript{\rm 3}McGill University,
    \textsuperscript{\rm 4}Mila - Québec AI Institute,
    \textsuperscript{\rm 5}Macao Polytechnic University,
    \textsuperscript{\rm 6}Peking University,
    \textsuperscript{\rm 7}Shanghai Artificial Intelligence Laboratory
}

\begin{document}

\maketitle

\begin{abstract}
Protein backbone generation models are often credited with exploring novel fold space based solely on low full-chain similarity to known proteins, yet this cannot distinguish a genuinely new fold from a novel assembly of known structural units. We first ask whether this granularity mismatch alone explains the reported rates, and introduce the Domain Retrieval Rate (DRR), the fraction of generated backbones for which any constituent domain matches a known domain in CATH S40. Applied to eight backbone generation models spanning diffusion and flow-matching paradigms, DRR finds locally alignable known structure in most outputs, while the fraction containing a substantially covered complete domain is considerably smaller and depends on the scoring convention. To calibrate what retrieval alone can achieve, we propose RetFold, a zero-training baseline that constructs backbones by retrieving CATH domains and refining inter-domain connections through geometry-based helix-linker optimization, at two orders of magnitude lower cost on CPU alone.
RetFold is recombined by construction, yet the conventional full-chain protocol retrieves it at only 20.0\%. We calibrate the three scores in use against this control and against a leave-one-topology-out negative control, and only one of the three supports reading a retrieval rate as a novelty estimate, and only at a threshold stricter than the conventional 0.5: query-normalized TM carries a length ceiling that suppresses long chains regardless of content, aligned-length TM retrieves 90.04\% of queries whose fold class has been removed from the searched set, and reference-normalized TM reaches a 5.38\% false positive rate only at 0.7. DRR removes the length ceiling but fails for the second of these reasons rather than the first.
Full-chain novelty is insufficient evidence of fold-level innovation, and novelty evaluation should report the calibration of the score it uses.
\end{abstract}

\section{Introduction}

Protein backbone generation models are increasingly evaluated for their ability to explore fold space beyond known proteins. Recent generators spanning diffusion~\cite{watson2023novo,yim2023se,ingraham2023illuminating} and flow matching~\cite{yim2023fast} report substantial fractions of outputs with no close full-chain match in a chosen structural database. Such results are often interpreted as evidence of previously unseen folds. This interpretation rests on two observations: generated backbones pass inverse-folding and structure-prediction validation~\cite{dauparas2022proteinmpnn,abramson2024accurate}, supporting in-silico designability, and they exhibit low full-chain similarity under a specified search protocol. We argue that these observations do not by themselves identify the structural scale at which novelty occurs. The gap between full-chain and domain-level retrievability instead exposes how strongly a novelty label depends on the evaluation protocol and on whether known structural units are considered separately.

\begin{figure}[t]
\centering
\includegraphics[width=0.48\textwidth]{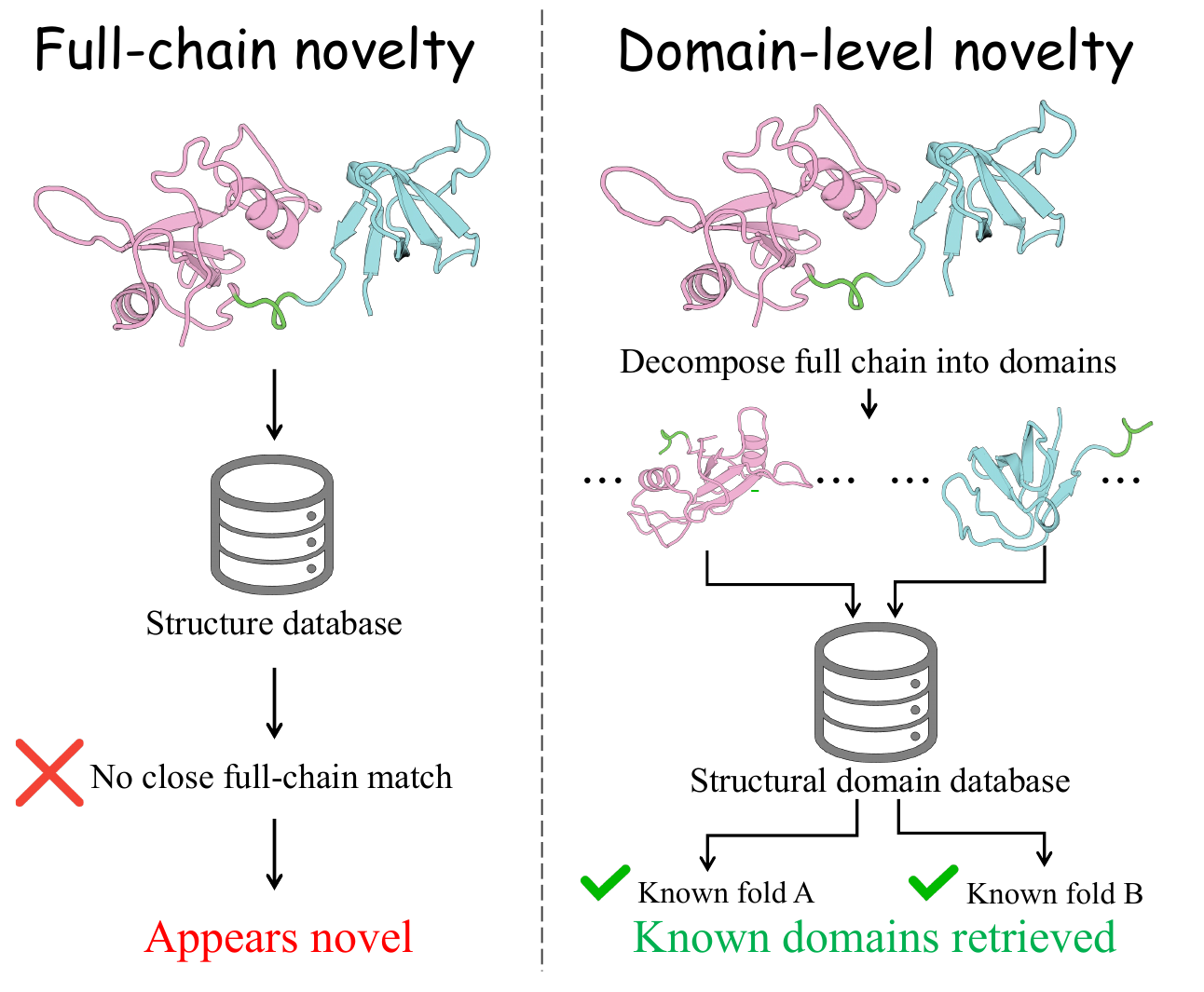}
\caption{
Contrasting full-chain and domain-level novelty. \textbf{Left:} The conventional approach compares a generated backbone as a monolithic unit. \textbf{Right:} Our DRR first decomposes the generated backbone into constituent domains and retrieves each against a structural domain database.
}
\label{fig:novelty_metric_comparsion}
\end{figure}

The core issue is a granularity mismatch. Full-chain comparison treats a generated backbone as a monolithic unit, yet natural proteins are modular: they are composed of folding domains that recombine across evolution~\cite{chothia1992one,orengo1997cath}. A structure that concatenates two retrievable domains in a previously unseen orientation can receive a low full-chain similarity score even though its components match entries in the reference database. This is combinatorial novelty relative to that database and protocol, rather than direct evidence of a new domain-level fold. Conflating these two scales can make recombination appear as fold-space expansion because full-chain comparison cannot distinguish rearrangement from invention. Figure~\ref{fig:novelty_metric_comparsion} illustrates this contrast: a backbone with no full-chain match may still decompose into domains that are individually retrievable under the same reference setting.
Physical plausibility is a separate confounder because unrealistic backbones can appear novel simply by matching no natural structure. We therefore report designability separately and evaluate retrieval on a fixed all-generated cohort.

\begin{figure*}[!t]
\centering
\includegraphics[width=\textwidth]{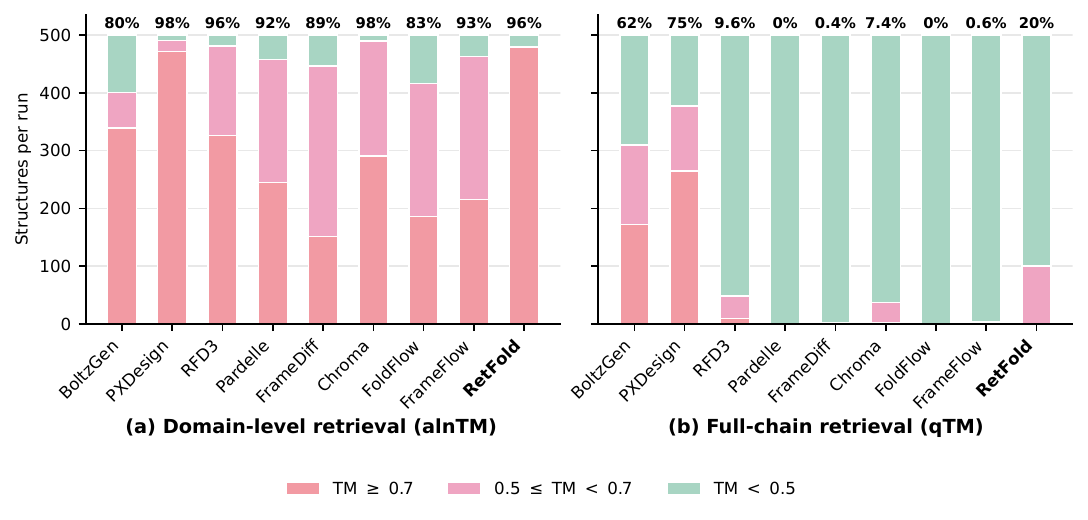}
\caption{Domain-level vs.\ full-chain retrievability under the unified Foldseek protocol and a fixed denominator of 500 per method. These two searches use CATH S40, sensitivity 9.5, and exact TM-score rescoring of the top 100 prefilter candidates.}
\label{fig:domain_vs_fullchain}
\end{figure*}

Those two issues motivate a domain-centric evaluation framework. We introduce Domain Retrieval Rate (DRR), a metric that decomposes generated backbones into structural domains and measures the proportion whose constituent domains match entries in CATH S40~\cite{orengo1997cath}. By comparing domain-level and full-chain retrievability, DRR quantifies the scale dependence of an operational novelty label. Across the eight generators we evaluate, aligned-length domain retrieval remains high even when full-chain retrieval varies widely (Figure~\ref{fig:domain_vs_fullchain}). This test is itself conditional on a score, however, and we show below that the aligned-length score it uses fails a specificity check that the full-chain score passes. A reference-normalized and coverage-controlled analysis provides a stricter boundary on how much of each reference domain is recovered.

To calibrate what retrieval alone can achieve, we propose RetFold, a zero-training baseline that constructs protein backbones by retrieving CATH domains and assembling them through geometry-based helix-linker optimization. RetFold requires no GPU training and no learned generative prior. Because its composition is fixed by construction, it supplies a reference point with a known answer: any criterion that fails to retrieve it is failing on a case where the correct label is not in doubt. If a training-free pipeline lands inside the retrieval envelope reported for trained generators, that envelope cannot by itself certify learned fold invention.
In summary, the contributions are as follows:
\begin{itemize}
\item We show that the retrieval scores underlying novelty claims in this area are reported without calibration, and calibrate all three against a positive control whose composition is known by construction and a leave-one-topology-out negative control.
\item Across eight representative models spanning diffusion and flow-matching
paradigms, we show that novelty labels vary substantially across operational
protocols, and that low full-chain retrievability coexists with high
domain-level retrieval under a score whose specificity we then quantify.
\item We construct RetFold, a backbone set assembled from unmodified CATH domains at 1.98\,s per structure on CPU. Its composition is fixed, so the correct answer under any retrieval criterion is known in advance; it lands inside the retrieval envelope reported for learned generators, while junction-resolved AF3 confidence shows what that envelope does not measure.
\end{itemize}

\section{Related Work}

\subsection{Protein Backbone Generation}

Generative models for protein backbones fall into a few families. Diffusion
models denoise residue-level coordinate frames: RFdiffusion performs
structure-conditioned denoising over SE(3) frames~\cite{watson2023novo}, FrameDiff casts backbone generation as
diffusion on residue rigid-body frames~\cite{yim2023se}, and Chroma pairs
graph-neural-network denoising with programmable conditioning for large-scale
sampling~\cite{ingraham2023illuminating}. Flow-matching models replace the
diffusion process with continuous normalizing flows on the SE(3) manifold to
improve sampling efficiency, as in FrameFlow~\cite{yim2023fast} and
FoldFlow~\cite{bose2024se3}. ProtPardelle extends generation to the
all-atom setting, with separately released unconditional backbone-only and all-atom sampling modes~\cite{chu2024all}. A separate line of work targets conditional
binder design rather than unconditional backbones: BoltzGen unifies generative
design with structure prediction to produce binders across biomolecular
modalities~\cite{stark2025boltzgen}, and PXDesign couples a diffusion generator
with confidence-based filtering for de novo binder design~\cite{team2025pxdesign}. We evaluate ProtPardelle in its unconditional backbone-only mode. For the primary task-matched backbone comparison, BoltzGen uses its released unconditional monomer specification, while PXDesign uses target-free inference with all condition masks set to zero; the latter is an inference ablation rather than a documented PXDesign application mode.
Rather than proposing a new generator, we introduce a retrieval-and-recombination baseline and evaluate both the baseline and existing generators under identical metrics, to check whether those metrics can truly distinguish learned generation from simple recombination.

\subsection{Fragment Assembly and Recombination}

Building structures from retrieved pieces long predates learned generation.
Rosetta \emph{ab initio} folding assembles candidate conformations from libraries
of short structural fragments drawn from known
proteins~\cite{simons1997assembly,rohl2004protein}. In protein engineering,
SCHEMA recombination composes chimeric proteins from structural blocks of known
parents, exploiting the observation that recombination preserves folded structure
when block boundaries respect contact density~\cite{voigt2002protein}. Both lines
establish that known structural units, recombined under geometric constraints,
yield foldable proteins that are not copies of any single parent. RetFold is
deliberately in this tradition, at domain rather than fragment granularity; we do
not claim it as a competitive design method, and its purpose is comparative.



\subsection{Representations and Domain Analysis}

Our approach reuses existing representations and tools for a different end:
calibrating novelty claims rather than predicting or annotating structures.
RetFold's retrieval combines ESM-2 sequence embeddings~\cite{lin2023esm2},
Foldseek's 3Di structural tokens~\cite{van2024fast}, and
ProstT5~\cite{heinzinger2024prostt5}. CATH organizes domains by
class, architecture, topology, and homologous superfamily; its non-redundant S40
subset (approximately 34{,}653 domains) serves as our map of known fold
space~\cite{cath}, and Merizo segments multi-domain chains into their constituent
structural units directly from backbone coordinates~\cite{lau2023merizo}.

\subsection{Backbone Quality and Novelty Evaluation}

Generated backbones are usually judged by a two-step designability test followed
by a novelty check. ProteinMPNN designs sequences conditioned on backbone
geometry~\cite{dauparas2022proteinmpnn}, and AlphaFold-family predictors check
whether those sequences fold back to the intended structure~\cite{jumper2021alphafold,abramson2024accurate}. Novelty is then
measured by full-chain alignment against known structures, typically via TM-score
or its query-normalized variant, with a backbone counted as novel when no
sufficiently similar full-chain match exists.
This protocol reliably catches near-exact structural reuse, but it has two blind
spots. It treats a multi-domain backbone as a single object, so it cannot separate
a genuinely new domain-level fold from a new arrangement of known domains. And when non-designable backbones remain in the novelty denominator, physically implausible structures can register as novel simply because nothing natural resembles them. We address both by decomposing retrievability across domain, local-alignment, and full-chain levels, and by reporting designability alongside novelty rather than mixing the two, so that non-designable structures neither inflate nor are silently excluded.

\section{Method}

\begin{figure*}[t]
\centering
\includegraphics[width=1\textwidth]{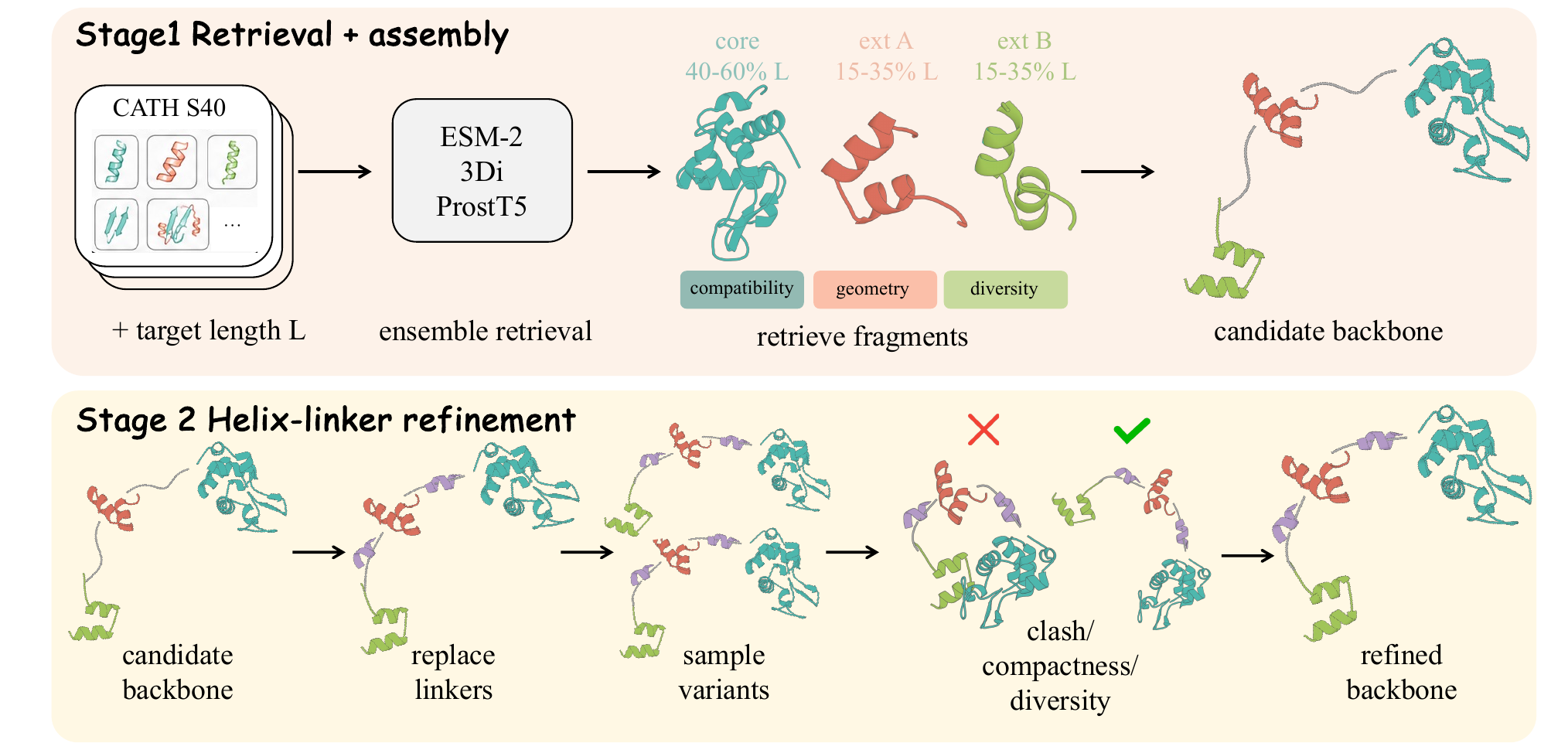}
\caption{Overview of RetFold. CATH S40 domains are embedded with complementary representations. RetFold retrieves one core fragment and two extension fragments under target-length constraints, assembles them into candidate full-chain backbones, generates helix-linker variants, and filters candidates by geometric validity and diversity.}
\label{fig:method_overview}
\end{figure*}

\subsection{Preliminaries}

\paragraph{Setup and notation.}
Let $\mathcal{G} = \{g_1, \ldots, g_N\}$ denote a set of generated backbones and $\mathcal{R}$ the non-redundant CATH S40 reference set. All retrieval decisions take the same form: a query is searched against $\mathcal{R}$, and is deemed \textit{retrievable} if its best hit attains a structural-similarity score at or above a threshold $\tau$. Unless stated otherwise we report a permissive $\tau = 0.5$, conventionally read as a same-fold criterion~\cite{xu2010significant}, and a high-confidence $\tau = 0.7$. Every rate below is therefore a statement about CATH S40 under a stated segmentation, normalization, threshold and coverage rule, not about absence from an exhaustive fold universe.

\paragraph{Conventional full-chain novelty assessment.}
Existing evaluations assess novelty at the full-chain level: each generated backbone $g$ is treated as a single query and aligned against $\mathcal{R}$ using the query-normalized TM-score $\mathrm{qTM}$. The full-chain retrieval indicator is
\begin{equation}
\mathbb{R}^{\mathrm{fc}}_\tau(g) = \mathbb{1}\!\left[ \max_{r \in \mathcal{R}} \mathrm{qTM}(g, r) \geq \tau \right],
\end{equation}
and the full-chain retrieval rate over $\mathcal{G}$ is $\mathrm{FC}(\tau) = \tfrac{1}{N}\sum_{g \in \mathcal{G}} \mathbb{R}^{\mathrm{fc}}_\tau(g)$. A backbone is counted as \textit{novel} when it is not retrievable.

\paragraph{The query-length ceiling.}
Because $\mathrm{qTM}$ divides the alignment sum by the query length and each summand is at most one, a query of length $L_q$ searched against a reference of length $L_r<L_q$ satisfies
\begin{equation}
\mathrm{qTM} \;\le\; \frac{N_{\mathrm{aligned}}}{L_q} \;\le\; \frac{L_r}{L_q}.
\label{eq:ceiling}
\end{equation}
Since CATH S40 contains domains rather than chains (median 132 residues), a chain of 500 residues must find a reference domain of at least 250 residues before $\mathrm{qTM}\geq0.5$ is attainable at all; only 13.4\% of the database qualifies (Appendix~\ref{app:calibration}). The ceiling is a property of the normalization and the reference database, not of the structure being scored. A domain-level view removes it by shortening the denominator, and a reference-normalized score removes it by replacing the denominator; we report both.

\subsection{Domain Retrieval Rate}

\paragraph{From whole chains to fold units.}
Since the query-length ceiling arises from treating each backbone as a single monolithic query, we also evaluate retrievability at the level of its constituent fold units. We decompose every generated backbone into structural domains with Merizo~\cite{lau2023merizo},
\begin{equation}
\mathrm{Decompose}(g) = \mathcal{D}(g) = \{d^1, d^2, \ldots, d^{M}\},
\end{equation}
where $M$ is the predicted domain count. This reframes the novelty question from \textit{does the whole chain resemble a known structure?} to \textit{do its fold units resemble known domains?}

\paragraph{Any-domain criterion and DRR.}
Each domain $d^j$ is retrieved against $\mathcal{R}$ using $\mathrm{alnTM}$, the TM-score computed over the locally aligned region and normalized by the aligned length. We aggregate to the backbone level with an \textit{any-domain} rule: a backbone is domain-retrievable if at least one of its domains matches a known CATH S40 domain,
\begin{equation}
\mathbb{R}^{\mathrm{dom}}_\tau(g) = \mathbb{1}\!\left[ \exists\, d^j \in \mathcal{D}(g),\ \exists\, r \in \mathcal{R}: \ \mathrm{alnTM}(d^j, r) \geq \tau \right],
\end{equation}
and the \textbf{Domain Retrieval Rate} is the fraction of backbones,
\begin{equation}
\mathrm{DRR}(\tau) = \frac{1}{N}\sum_{g \in \mathcal{G}} \mathbb{R}^{\mathrm{dom}}_\tau(g).
\end{equation}
The any-domain rule is a deliberately \textit{lenient} test of structural reuse: it certifies that a backbone contains at least one known fold unit, not that all of its domains are known. Its force therefore comes not from its own magnitude but from the contrast with full-chain novelty. If backbones that full-chain metrics deem novel nonetheless contain retrievable domains under even this permissive criterion, then at least part of their apparent novelty may reflect known fold units rather than full-chain fold invention. We report the stricter all-domain rate, which requires every domain to be retrievable, in Table~\ref{tab:all_domain}.


\paragraph{Reference-normalized, coverage-controlled retrieval.}
Because alnTM deliberately credits high-quality local correspondence, we test whether a match substantially covers a complete CATH reference domain. A strict hit requires both a target-normalized $\mathrm{tTM}\geq0.5$ and target coverage $\mathrm{tcov}\geq0.7$. This is a robustness boundary, not an alternative DRR definition: DRR measures reuse of an alignable known unit, whereas the strict criterion asks whether that unit is \emph{complete} enough to support a known-fold interpretation. Appendix~\ref{app:retrievability} gives the formal definition and strict any/all-domain results.

\paragraph{The domain--full-chain gap.}
We summarise the two rates by their difference,
\begin{equation}
\Delta(\tau) = \mathrm{DRR}(\tau) - \mathrm{FC}(\tau),
\end{equation}
which contrasts what a domain-level, aligned-length protocol recovers with what a full-chain, query-normalized protocol recovers. The two differ in granularity and normalization, so $\Delta$ should be read as a comparison between two conventions rather than as a decomposition of novelty into components.
We report the value $\Delta$ takes when neither protocol has a true match to find below, which bounds how much of an observed gap can be attributed to structural content.


\paragraph{Boundary-agnostic robustness check.}
We also report the same score applied to the intact chain rather than to its
predicted domains. This fires whenever an alignable known fragment exists,
independent of domain boundaries.

\subsection{RetFold: Retrieval-and-Assembly Baseline}

To assess how much of the reported retrieval profile can be reproduced without learning, we propose \textbf{RetFold}, a zero-training baseline that constructs backbones by retrieving domains and assembling them into multi-fragment structures. RetFold does not learn a generative prior; its role is diagnostic. Its composition is fixed by construction, so the correct answer for any retrieval criterion applied to it is known in advance. If a training-free pipeline lands inside the domain-level and full-chain retrieval envelope reported for learned generators, then that envelope cannot by itself certify that a generator has invented new folds.

Figure~\ref{fig:method_overview} shows that RetFold proceeds in two stages. \textbf{Stage~1} (ensemble-guided retrieval and assembly) retrieves three CATH S40 fragments (one core fragment at 40--60\% of target length $L$ and two extensions at 15--35\% each) using an ensemble embedding that combines ESM-2, Foldseek 3Di, and ProstT5. Fragments are selected by embedding compatibility, diversity, and geometric constraints, then assembled into candidate backbones. \textbf{Stage~2} (helix-linker refinement) replaces flexible inter-fragment linkers with idealized $\alpha$-helical connections, sampling orientation and rotational phase to generate geometric variants. Variants are filtered by steric clash, compactness, and diversity; the selected final variants are then evaluated with ProteinMPNN ($K{=}8$) and AlphaFold3 ($\mathrm{pLDDT} \geq 70$). We note that domain-level foldability is largely by construction, since Stage~1 fragments are real CATH domains; the non-trivial designability burden falls on the Stage~2 inter-fragment junctions, where the helix-linker refinement operates.

\section{Experiments}

We evaluate eight protein backbone generation models: RFDiffusion~\cite{watson2023novo}, FrameDiff~\cite{yim2023se}, Chroma~\cite{ingraham2023illuminating}, FoldFlow~\cite{bose2024se3}, FrameFlow~\cite{yim2023fast}, ProtPardelle~\cite{chu2024all}, BoltzGen~\cite{stark2025boltzgen}, and PXDesign~\cite{team2025pxdesign}. The primary comparison uses a fixed planned denominator of 500 backbones per method, with 100 at each length $L \in \{100, 200, 300, 400, 500\}$; two unavailable FrameDiff outputs are counted as unretrieved. ProtPardelle uses its unconditional backbone-only mode, while BoltzGen uses released length-specific monomer specifications. PXDesign is run through a target-free inference ablation in which all residues are design tokens and structural-condition masks are zeroed.
For these two task-matched cohorts, ProteinMPNN generates eight candidate sequences per backbone and one is forwarded to AF3, whereas the other methods use best-of-$K$; an output passes when its mean pLDDT is at least 70. Comparisons conditioned on AF3 outcome therefore favour the best-of-$K$ cohorts, which is one reason we use the all-generated denominator for the primary analysis.
The primary DRR comparison includes all generated backbones regardless of AF3 outcome. For RetFold, we retain 500 final backbones, with 100 at each target length, after geometric-validity, geometric-score, and diversity filtering of the Stage~2 candidate pool. DRR is computed over these 500 selected final outputs. We therefore use RetFold as a retrieval-based calibration baseline and do not interpret its retained-output rate as directly comparable to the raw sampling yield of learned generators.

\subsection{The Domain \& Full-chain gap}
\label{sec:res-gap}

\paragraph{Domain-level retrieval and full-chain novelty.} Under the unified evaluation, DRR at alnTM${\geq}0.5$ ranges from 80.2\% for BoltzGen to 98.2\% for PXDesign; RetFold achieves 96.0\%. Thus, most outputs from every evaluated generator contain at least one region alignable to a CATH S40 domain under this protocol. Full-chain retrievability at qTM${\geq}0.5$ is far lower for most models: 75.4\% for PXDesign, 62.0\% for BoltzGen, 20.0\% for RetFold, 9.6\% for RFDiffusion, 7.4\% for Chroma, and at most 0.6\% for the remaining models. The two task-matched binder systems retain high full-chain retrievability, while the six unconditional generators fall between 0.0\% and 9.6\%. The all-domain criterion in Table~\ref{tab:all_domain} narrows the picture: requiring \emph{every} predicted domain to match leaves 28.4--69.0\% for the six unconditional generators, so the aligned-length any-domain rate is carried in part by backbones with only one matching unit.

\begin{table}[ht]
\centering
\small
\setlength{\tabcolsep}{4.6mm}
\caption{Any- vs.\ all-domain retrieval rates (\%) under the primary top-100, all-generated protocol. Any-domain counts a backbone if at least one domain matches CATH S40; all-domain requires all domains to match. Rates use aligned-length TM, whose specificity is calibrated in Table~\ref{tab:app_loto}; they bound local alignability rather than complete domain reuse.}
\begin{tabular}{lrrrr}
\toprule
 & \multicolumn{2}{c}{$\tau = 0.5$} & \multicolumn{2}{c}{$\tau = 0.7$} \\
\cmidrule(lr){2-3} \cmidrule(lr){4-5}
Method & Any & All & Any & All \\
\midrule
RFDiffusion  & 96.2 & 60.6 & 65.2 & 20.6 \\
FrameDiff    & 89.4 & 47.8 & 30.2 & 10.0 \\
Chroma       & 98.0 & 69.0 & 58.2 & 12.8 \\
FoldFlow     & 83.2 & 28.4 & 37.2 & 7.2 \\
FrameFlow    & 92.6 & 49.6 & 43.0 & 10.2 \\
ProtPardelle & 91.6 & 47.0 & 49.0 & 14.0 \\
BoltzGen   & 80.2 & 64.4 & 67.8 & 51.4 \\
PXDesign  & 98.2 & 87.6 & 94.4 & 82.0 \\
\midrule
RetFold      & 96.0 & 83.8 & 95.8 & 81.8 \\
\bottomrule
\end{tabular}
\label{tab:all_domain}
\end{table}

\paragraph{How the rate depends on the protocol.} The rate is conditional on four choices, and varies substantially across all. 
\begin{itemize}
\item \textbf{\emph{Reference normalization and coverage.}} Reference-normalized, coverage-controlled retrieval tests the stronger requirement that a match substantially cover a complete reference domain (Appendix~\ref{app:retrievability}, Figure~\ref{fig:reference_normalized_drr}). Strict any-domain retrieval is 95.4\% for RetFold, 86.4\% for PXDesign, and 62.0\% for BoltzGen, versus 41.8\% for both FoldFlow and FrameFlow and 31.6--50.4\% across the remaining learned generators. Aligned-length DRR thus identifies alignable structure under a score whose specificity is quantified below, while the stricter protocol narrows the interpretation of permissive DRR.
\item \textbf{\emph{Segmentation.}} Intact-chain aligned-length retrieval and an independent Chainsaw analysis give boundary-agnostic and alternative-segmentation checks (Appendices~\ref{app:retrievability} and~\ref{app:segmentation}). Under Chainsaw, DRR at alnTM${\geq}0.5$ ranges from 39.2\% to 99.8\%; most methods stay above 50\%, though FoldFlow loses majority retrieval. RetFold gives the one case where the correct segmentation is known: its chains contain three fragments by construction, and both tools report fewer (2.43 and 2.27), so segmentation loss is present even where the units are unmodified reference domains. Merizo is retained for the primary analysis because it defines the units used by DRR.
\item \textbf{\emph{Quality conditioning.}} The separation persists when the
analysis is restricted to AF3-passed structures (Appendix~\ref{app:retrievability},
Table~\ref{tab:app_af3_ablation}). Domain-level retrieval shifts by at most 1.5
points for six of nine methods; full-chain retrieval rises for the low-pass-rate
cohorts, most notably Chroma (7.4\% to 26.3\%, 99 passed), so the unfiltered
denominator is the more conservative choice for the gap.
\item \textbf{\emph{Target length.}} Across learned generators DRR exceeds FC in most model--length cells, while FC generally declines with length (Appendix~\ref{app:retrievability}, Figure~\ref{fig:app_per_length_all}). Domain count correlates strongly with length, but a model- and length-adjusted analysis does not support an independent decrease in FC with the number of Merizo domains; we therefore read these cells as descriptive trends, not a universal domain-count mechanism.
\end{itemize}

\paragraph{What the aligned-length rates support.} Because these rates are conditional on a score and a threshold, we calibrate all three scores in both directions (Appendix~\ref{app:calibration}). On the RetFold control, whose composition is known by construction, a reference-normalized criterion is saturated (500/500 at $\mathrm{tTM}\geq0.9$) while the conventional full-chain $\mathrm{qTM}$ protocol retrieves 20.0\%. On a leave-one-topology-out negative control, $\mathrm{alnTM}\geq0.5$ retrieves 90.04\% of queries whose fold class has been removed from the searched set, against 29.97\% for $\mathrm{tTM}$ and 21.06\% for $\mathrm{qTM}$. Aligned-length rates therefore bound local alignability rather than complete domain reuse (Table~\ref{tab:app_loto}).

\begin{table}[ht]
\centering
\caption{Leave-one-topology-out calibration over 34{,}653 CATH S40 queries.
FPR is the fraction retrieved after candidates from the query's own topology are
discarded; a score with no discriminative power would reach 100. All rates use
domain-sized queries and do not describe behaviour on multi-domain chains: on the
RetFold control $\mathrm{qTM}$ retrieves 20.0\%
(Appendix~\ref{app:calibration}). AUC does not depend on $\tau$.}
\small
\setlength{\tabcolsep}{3.2mm}
\begin{tabular}{lrrrrr}
\toprule
 & \multicolumn{2}{c}{$\tau=0.5$} & \multicolumn{2}{c}{$\tau=0.7$} & \\
\cmidrule(lr){2-3}\cmidrule(lr){4-5}
Score & FPR & Sens. & FPR & Sens. & AUC \\
\midrule
$\mathrm{alnTM}$ & 90.04 & 97.48 & 50.63 & 89.39 & 0.809 \\
$\mathrm{qTM}$   & 21.06 & 92.37 & 3.79  & 73.24 & 0.936 \\
$\mathrm{tTM}$   & 29.97 & 93.85 & 5.38  & 77.48 & 0.926 \\
\bottomrule
\end{tabular}
\label{tab:app_loto}
\end{table}

\paragraph{What $\Delta$ measures.} $\Delta$ subtracts a query-normalized
full-chain rate from an aligned-length domain rate, so it inherits the
false-positive behaviour of the aligned-length score on its positive term: a set
of domain queries with no true match in the searched set is still assigned
$\mathrm{DRR}=90.04\%$ (Table~\ref{tab:app_loto}). We use that rate alone as the
reference, since Eq.~\ref{eq:ceiling} drives the $\mathrm{qTM}$ null toward zero
on chain-sized queries; using the domain-sized $\mathrm{qTM}$ null of 21.06\%
instead would place the reference at 68.98 and every observed value above it.
The observed $\Delta$ range from 18.2 points (BoltzGen) to 92.0 points
(FrameFlow): no method exceeds the reference by more than 2.0 points, and six of
the nine fall below it. RetFold, whose composition is entirely known and entirely
recombined, has the lowest $\Delta$ of the seven cohorts run in a documented
mode; BoltzGen (18.2) and PXDesign (22.8) fall lower still, but both are
target-free ablations. If $\Delta$ tracked
recombination the ordering would be reversed. Its FC is elevated (20.0\%) because
its core fragment is a verbatim reference domain spanning 40--60\% of the chain,
so $N_{\mathrm{aligned}}/L_q$ straddles the 0.5 threshold, whereas the learned
generators align roughly one Merizo domain (122--138 residues) against chains of
up to 500. Under this upper-bound estimate of the aligned-length false positive
rate, $\Delta$ does not carry information about structural content at the
resolution at which it has been used.

\begin{boxA}
\noindent\textbf{Takeaway.}
Apparent full-chain novelty coexists with high domain retrievability. Measured
against the value $\Delta$ takes when no true match exists, no method exceeds the
reference by more than 2.0 points and six of the nine fall below it. The gap is a
property of the two normalization conventions, not of the structures.
\end{boxA}

\subsection{A zero-training baseline exposes the resolution limit}
\label{sec:res-retfold}

A zero-training retrieval-and-assembly pipeline reaches the retrieval profile reported for learned generators under the conventional protocol. RetFold reaches DRR 96.0\% at alnTM${\geq}0.5$ while its full-chain retrievability stays at 20.0\% (qTM${\geq}0.5$), a 76.0-point domain--full-chain gap obtained through explicit retrieval and geometry-based assembly. The value of this comparison is that RetFold's composition is not in question: every chain contains complete CATH domains by construction and the conventional protocol nonetheless assigns it 20.0\%.

The match is not limited to the headline gap; RetFold sits inside the
reported range on the other axes as well.
\begin{itemize}
\item \textbf{\emph{Per-length behavior.}} RetFold's domain-level retrieval is stable at lengths L200 and above (Appendix~\ref{app:length-cost}, Table~\ref{tab:app_retfold_length}). Its lower L100 rate at $\tau{=}0.7$ reflects the limited decomposition of short, often single-domain outputs, for which Merizo returns no separable units.
\item \textbf{\emph{Diversity and coverage.}} RetFold reaches mean pairwise qTM
of 0.18 and spans 82 CATH topologies and 143 superfamilies. At a common rarefied sample its topology coverage overlaps that of RFDiffusion
(Appendix~\ref{app:diversity}).
\item \textbf{\emph{Computational cost.}} The workflow takes 1.98\,s per backbone
on CPU, about 110$\times$ faster than the measured RFDiffusion run
(Appendix~\ref{app:length-cost}, Table~\ref{tab:app_timing}). Timings exclude
shared downstream validation and are descriptive rather than hardware-normalized.
\end{itemize}



Reproducing the retrieval \emph{statistics} does not imply solving inter-domain design. Whole-chain average pLDDT masks a localized weakness: among RetFold's 466 AF3-passed backbones, inserted helix-linker residues have mean pLDDT 43.1 versus 79.1 for retrieved-fragment residues (paired mean difference $-36.0$; 95\% bootstrap CI $[-36.8,-35.2]$), and mean inter-fragment PAE is 24.9\,\AA. Figure~\ref{fig:app_junction_confidence} localizes the confidence drop to the fragment--linker boundaries. RetFold should therefore be read as a calibration baseline for retrieval statistics, not as evidence that geometric refinement alone solves inter-domain designability.


\paragraph{A calibrated criterion separates them.} Applying the
reference-normalized criterion to full-chain queries
(Table~\ref{tab:app_ttm_threshold}) gives the comparison the conventional
protocol cannot make. At $\tau{=}0.5$ the recombined control is retrieved at
100.0\% against 0.0--50.0\% for the six unconditional generators. Raising the
threshold to 0.7, which reduces the false positive rate to 5.38\%,
leaves the control unchanged at 100.0\% while Chroma falls 47.0 points and
RFDiffusion 36.2. A chain assembled from complete reference domains scores near
1.0 and is threshold-invariant over $[0.5,0.9]$; the learned generators instead
concentrate in $[0.5,0.7)$, falling to 3.0\% and 4.4\% at $\tau{=}0.7$. Their
relationship to known domains is neither complete reuse nor absence.

The two cohorts that clear the background by a wide margin, PXDesign and
BoltzGen, are precisely the two run outside their documented inference mode, so
we do not read this as a property of conditional binder design in general.

The unretrieved fraction is not a count of new folds. A backbone can fail this
criterion because it contains a genuinely unseen fold, or because it contains a
distorted version of a known one that no longer covers the reference domain. Our
data do not separate these, and the low AF3 pass rates of several of these
cohorts (14--28\% for FrameDiff, Chroma and ProtPardelle) suggest the second case
is not rare. Distinguishing them requires a designability-stratified analysis
that we leave to future work.

\begin{table}[ht]
\centering
\caption{Full-chain target-normalized retrieval over the all-generated cohort
(500 backbones per method). The leave-one-topology-out false positive rate at
$\tau{=}0.7$ is 5.38\%; it is measured on domain-sized queries and bounds that
query population rather than transferring directly to generated chains.}
\small
\setlength{\tabcolsep}{4mm}
\begin{tabular}{lrrr}
\toprule
Method & $\mathrm{tTM}\geq0.5$ & $\mathrm{tTM}\geq0.7$ & Drop \\
\midrule
RetFold      & 100.0 & 100.0 & 0.0 \\
\midrule
PXDesign     & 95.2 & 88.4 & 6.8 \\
BoltzGen     & 73.8 & 47.2 & 26.6 \\
Chroma       & 50.0 & 3.0  & 47.0 \\
RFDiffusion  & 40.6 & 4.4  & 36.2 \\
FrameFlow    & 0.8  & 0.0  & 0.8 \\
FrameDiff    & 0.2  & 0.0  & 0.2 \\
ProtPardelle & 0.0  & 0.0  & 0.0 \\
FoldFlow     & 0.0  & 0.0  & 0.0 \\
\bottomrule
\end{tabular}
\label{tab:app_ttm_threshold}
\end{table}

\begin{boxA}
\noindent\textbf{Takeaway.}
A construction that is recombined by definition sits inside the DRR--FC envelope
reported for learned generators, at two orders of magnitude lower cost. The
envelope does not measure what it is taken to measure. A criterion
calibrated on both controls does separate the two, and gives a different picture
of how much known structure the generators reuse.
\end{boxA}

\subsection{The designability screen passes near-copies}

The same calibration question applies to the designability screen. We retrieved
binders for five protein--protein complexes from PDB100 without homolog
exclusion, so the database contains near-identical complexes for several targets
(1ct4\_EI for 3sgb\_EI at 0.146\,\AA). Each retrieved binder was superposed onto
the native interface, redesigned with ProteinMPNN, and evaluated with AlphaFold3.

Under the standard screen ($\mathrm{iPTM}\geq0.6$, $\mathrm{pLDDT}\geq70$), 176
of 256 designs pass, and the pass rate declines monotonically with retrieval
RMSD, from 100\% at 0.15\,\AA\ to 25\% at 0.65\,\AA\ (Appendix~\ref{app:binder}).
The screen does not register that these binders are retrieved near-copies; what
it tracks is how close the retrieved template happens to be. A designability rate
reported without a matched trivial baseline therefore does not by itself
establish that a pipeline has designed anything.

\section{Conclusion and Discussion}

In this study, we show that low full-chain similarity does not by itself establish new fold discovery. Under the conventional protocol, most generated backbones contain a retrievable known domain even when the full chain looks novel, and a zero-training retrieval baseline reproduces the same profile. Calibrating the underlying scores bounds this reading: aligned-length retrieval recovers 90.04\% of queries whose fold class has been removed from the reference set, so it evidences local alignability rather than complete domain reuse. Under a reference-normalized criterion calibrated on both a positive and a negative control, the recombined baseline and the six unconditional generators fall on
opposite sides of the threshold, which the conventional protocol does not show. We hope this motivates novelty evaluation that reports the calibration of its score alongside the rate.

However, our analysis is bounded by the coverage of CATH S40, which is a non-exhaustive snapshot of characterized domain space and will continue to expand as structural databases grow. The reported retrieval and novelty rates should be interpreted relative to this reference rather than as absolute estimates of all known or unknown folds.

\bibliography{aaai2027}

\clearpage

\appendix

\setcounter{secnumdepth}{1}

\section{Retrieval-Based Binder Design}
\label{app:binder}

Beyond backbone generation, we test whether retrieval extends to functional protein design. This experiment is exploratory and is not used to support the aggregate RetFold claims in the main text.

\paragraph{Pipeline.} For each target complex, we retrieve structurally similar binder candidates from PDB100 using Foldseek with the full-chain query mode. The retrieved binder is superposed onto the target binding site via Kabsch alignment (minimizing C$\alpha$ RMSD between the retrieved binder and the native binder in the reference complex). The superposed binder chain is then redesigned with ProteinMPNN ($K{=}8$ sequences, target chain fixed as context) and validated with AlphaFold3 multimer prediction (iPTM${\geq}0.6$, pLDDT${\geq}70$).

\begin{figure*}[t]
\centering
\includegraphics[width=\textwidth]{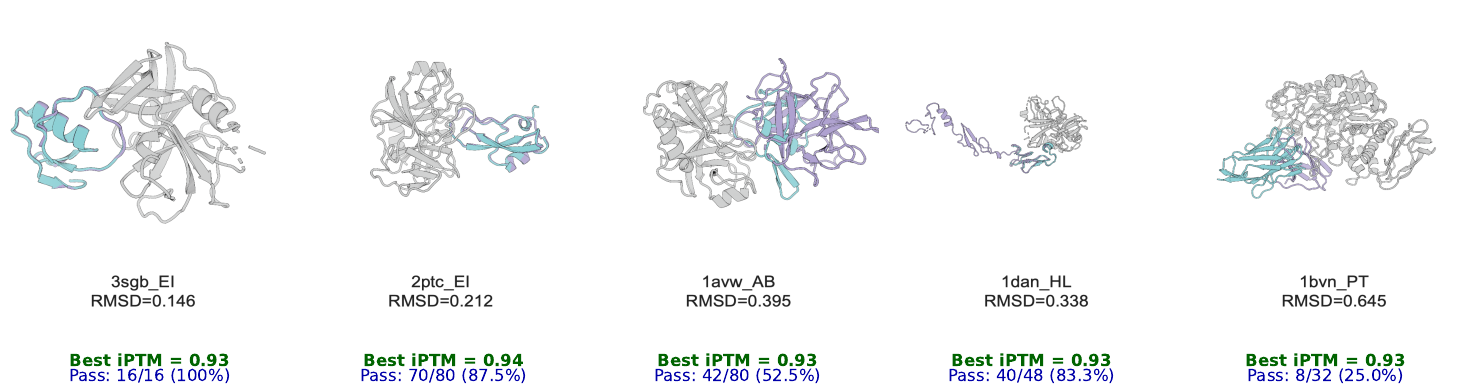}
\caption{Retrieval-based binder design for all five targets. Target chains are shown in gray; ground-truth binders from the original PDB complexes in purple; retrieved binders (superposed via Kabsch alignment) in cyan. Each panel reports the retrieval RMSD, best iPTM, and number of passing designs. The fifth target (1bvn\_PT, RMSD$=$0.645~\AA) achieves the lowest pass rate (25.0\%).}
\label{fig:app_binder}
\end{figure*}

\paragraph{Results.} Table~\ref{tab:app_cases} reports all five targets. Across 256 designed complexes, 176 (68.8\%) pass validation, with all five targets yielding at least one successful binder (best iPTM 0.93--0.94). All cases with recorded geometry achieve zero steric clashes after superposition and more than 45 interface contacts. Figure~\ref{fig:app_binder} superposes each retrieved binder (cyan) onto the ground-truth binding partner (purple) against the fixed target (gray). For 3sgb\_EI (RMSD$=$0.15\,\AA), the retrieved and ground-truth binders overlap almost perfectly and all 16 designs pass. As retrieval RMSD grows the two structures visibly separate and the pass rate declines monotonically, down to 25.0\% for the most distant target 1bvn\_PT (0.65\,\AA).

\paragraph{Quality metrics.} \textit{Superposition RMSD}: C$\alpha$ RMSD between the retrieved binder and the native binder after Kabsch alignment. \textit{Clashes}: backbone atom pairs between binder and target closer than 2.5~\AA\ after superposition. \textit{Interface contacts}: inter-chain C$\beta$ pairs within 8~\AA. \textit{Foldseek probability}: retrieval confidence.

\paragraph{Limitations of the case study.} These five targets represent well-characterized protein--protein interfaces with close structural homologs in PDB100, and no homolog-exclusion filter was applied to the retrieval database; the retrieved templates are in several cases near-identical complexes (e.g.\ 1ct4\_EI for 3sgb\_EI at 0.15\,\AA). The observed RMSD--success gradient should therefore be read as descriptive of this leakage-permissive setting rather than as evidence about the relative merits of retrieval and learned priors in general. The pipeline's success depends on database coverage: for binding interfaces without close structural analogs, retrieval-based design would fail. This contrasts with generative approaches (e.g., RFDiffusion hotspot conditioning) that can in principle design binders for novel interfaces, albeit with lower and less predictable success rates.

\begin{table*}[t]
\centering
\caption{Retrieval-based binder design across five protein--protein complexes. Superposition RMSD is computed over binder C$\alpha$ atoms after Kabsch alignment to the native binding pose. Pass: iPTM${\geq}0.6$ and pLDDT${\geq}70$ under AlphaFold3 multimer prediction. Foldseek retrieval probability was 1.0 for all recorded cases. The 1dan\_HL and 1bvn\_PT rows use the best candidates recorded in the assembly manifest.}
\small
\setlength{\tabcolsep}{3.2mm}
\begin{tabular}{llrrrrr}
\toprule
Target & Retrieved source & Superposition RMSD (\AA) & Clashes & Interface contacts & \#Passed/\#Designs & Pass rate \\
\midrule
3sgb\_EI & 1ct4\_EI & 0.146 & 0 & 51 & 16/16 & 100.0\% \\
2ptc\_EI & 2fi5\_EI & 0.212 & 0 & 45 & 70/80 & 87.5\% \\
1dan\_HL & 4x8u\_HL & 0.338 & 0 & 47 & 40/48 & 83.3\% \\
1avw\_AB & 1z7k\_AB & 0.395 & 0 & 52 & 42/80 & 52.5\% \\
1bvn\_PT & 1kxq\_AH & 0.645 & 0 & 58 & 8/32 & 25.0\% \\
\midrule
\textbf{Overall} & --- & --- & --- & --- & \textbf{176/256} & \textbf{68.8\%} \\
\bottomrule
\end{tabular}
\label{tab:app_cases}
\end{table*}

\section{Experimental and Metric Details}
\label{app:details}

This appendix collects what is needed to reproduce the analysis: the settings shared across all methods, the per-model generation configurations, the two algorithms, and RetFold's implementation constants.

\subsection{Evaluation Settings}

Table~\ref{tab:app_settings} lists the settings common to the generation models and to RetFold. Two choices deserve emphasis. First, the novelty denominator is the full generated set rather than the AF3-passed subset, so that non-designable structures neither inflate the apparent novelty nor are silently removed from it; the ablation in Appendix~\ref{app:retrievability} shows this choice does not drive the result. Second, domain-level and full-chain retrieval use different TM normalizations, an asymmetry quantified in Appendix~\ref{app:calibration}.

\begin{table*}[t]
\centering
\caption{Common experimental settings and retrieval thresholds used for the DRR and RetFold analyses. DRR is computed over all generated structures; AF3 pass rate is reported separately as a quality indicator.}
\small
\setlength{\tabcolsep}{1mm}
\begin{tabular}{ll}
\toprule
Component & Setting \\
\midrule
Generation models & RFDiffusion, FrameDiff, Chroma, FoldFlow, FrameFlow, ProtPardelle, BoltzGen, PXDesign \\
Generated backbones & 500 per method; 100 each at L100, L200, L300, L400, and L500 \\
Sequence design & ProteinMPNN with $K=8$ candidate sequences per backbone \\
Designability assessment & AF3 pLDDT $\geq 70$; one ProteinMPNN sequence per backbone for BoltzGen/PXDesign, and best-of-$K$ for others \\
Novelty denominator & $\mathcal{G}$, all generated backbones for each method (not restricted to AF3-passed) \\
Domain segmentation & Merizo applied to all generated structures before domain-level retrieval \\
Reference database & CATH S40 Foldseek database (34{,}653 non-redundant domains) \\
Retrieval levels & Domain-level alnTM (Merizo-decomposed domains vs.\ CATH S40) and full-chain qTM \\
Thresholds & TM $\geq 0.5$ (permissive) and $\geq 0.7$ (high-confidence) \\
RetFold output set & 500 selected final backbones; 100 per target length after geometric-validity and diversity filtering \\
\bottomrule
\end{tabular}
\label{tab:app_settings}
\end{table*}

\paragraph{Generation model configurations.} RFDiffusion uses the SE(3) diffusion framework with 200 denoising steps on 1 GPU. FrameDiff uses the SE(3) diffusion framework on 1 GPU. Chroma uses a diffusion model with programmable conditioning on 1 GPU. FoldFlow uses the published \texttt{ff2\_base.pth} checkpoint and FrameFlow uses the published \texttt{published.ckpt} checkpoint; both use 100 sampling steps, with one GPU assigned to each target-length job. ProtPardelle was not run in its all-atom codesign mode: we invoked \texttt{draw\_samples.py --type backbone} with \texttt{configs/uncond\_sampling.yml}, the released backbone checkpoint, 500 denoising steps, and 100 samples at each requested length; five single-GPU length jobs were scheduled across four GPUs. BoltzGen used the released length-benchmark monomer specifications, each containing a single protein entity of length 100--500, with protocol \texttt{protein-anything}, the \texttt{boltzgen1\_diverse} and \texttt{boltzgen1\_adherence} design checkpoints, and \texttt{--steps design}; generation used three GPUs, and its internal inverse-folding and refolding stages were skipped so that all methods used the shared ProteinMPNN--AF3 screen. PXDesign used one GPU with the released \texttt{pxdesign\_v0.1.0} checkpoint and 400-step sampler through the lower-level \texttt{infer} interface. Each input JSON contained one generated protein entity and no \texttt{condition} entry, producing a single chain with all condition masks zero; we therefore report this as a target-free inference ablation. We generated 100 structures per length and applied the shared ProteinMPNN--AF3 screen.

\paragraph{RetFold configuration.} RetFold was run entirely on CPU without GPU requirements. The CATH S40 domain embedding library (ESM-2, 3Di, ProstT5 representations for all 34{,}653 domains) was precomputed once and stored on disk. At generation time, RetFold loads the precomputed embeddings and performs retrieval, assembly, and refinement. The reported per-backbone times reflect only the retrieval-assembly-refinement compute; embedding precomputation is a one-time cost amortized over all experiments.

\subsection{DRR Evaluation Procedure}

Algorithm~\ref{alg:drr} gives the procedure used to compute DRR and FC. Each generated backbone is searched twice against the same reference set, once as a collection of Merizo domains and once as an intact chain, so that the two rates differ in how the query is presented rather than in what it is compared against. All generated structures enter the retrieval denominator; designability is assessed separately and reported as a quality indicator.

\begin{algorithm}[t]
\caption{DRR evaluation procedure}
\label{alg:drr}
\begin{algorithmic}[1]
\REQUIRE Generated backbones $\mathcal{G}$, reference database $\mathcal{R}$, threshold $\tau$
\FOR{each generated backbone $g_i \in \mathcal{G}$}
    \STATE Segment $g_i$ into Merizo domains $\mathcal{D}_i$
    \STATE Search each $d_i^j \in \mathcal{D}_i$ against $\mathcal{R}$ to compute $\mathbb{R}^{\mathrm{dom}}_\tau(g_i)$
    \STATE Search $g_i$ as an intact query to compute $\mathbb{R}^{\mathrm{fc}}_\tau(g_i)$
\ENDFOR
\STATE Compute $\mathrm{DRR}(\tau)$ and $\mathrm{FC}(\tau)$ over all $|\mathcal{G}|$ structures
\STATE Return $\mathrm{DRR}$ and $\mathrm{FC}$
\end{algorithmic}
\end{algorithm}

\subsection{RetFold Retrieval-and-Assembly Procedure}

Algorithm~\ref{alg:retfold} summarizes RetFold. The method is intentionally non-generative in the learning sense: it retrieves known CATH domains, assembles them into full-chain arrangements that do not occur in nature, and resolves the resulting inter-fragment junctions geometrically rather than through a learned prior. No step involves gradient-based training, and the only fitted quantities are the precomputed embeddings of the reference domains.

\begin{algorithm}[t]
\caption{RetFold retrieval-and-assembly baseline}
\label{alg:retfold}
\begin{algorithmic}[1]
\REQUIRE Target length $L$, CATH S40 domain library, ensemble embeddings
\STATE Select one core fragment with length 40--60\% of $L$
\STATE Select two extension fragments with length 15--35\% of $L$
\STATE Rank candidate fragment triples by ensemble embedding compatibility
\STATE Filter triples by CATH hierarchy distance and inter-extension similarity
\STATE Assemble each accepted core-extension triple into an initial backbone
\FOR{each initial backbone}
    \STATE Sample idealized $\alpha$-helix linkers of length 6--12 residues
    \STATE Randomize helix axis direction and rotational phase to generate linker variants
    \STATE Reject variants with backbone clashes or poor compactness
    \STATE Select diverse candidates by radius of gyration, contact count, and centroid distance bins
\ENDFOR
\STATE Run ProteinMPNN and AlphaFold3 validation on selected Stage 2 backbones
\STATE Return RetFold backbones for diversity, coverage, and cost analysis
\end{algorithmic}
\end{algorithm}

\subsection{RetFold Implementation Parameters}

\paragraph{Stage 1 (Ensemble Retrieval and Assembly).} Fragment length constraints: core domain 40--60\% of target length $L$, extension fragments 15--35\% of $L$ each. Ensemble embedding weights: 23\% for ESM-2, 38.5\% for Foldseek 3Di k-mer encoding, 38.5\% for ProstT5. Embedding compatibility window: $[0.45, 0.85]$ in cosine similarity. CATH hierarchy distance between fragments $\geq 2$. Maximum inter-extension similarity $< 0.75$.

\paragraph{Stage 2 (Helix-Linker Refinement).} Idealized $\alpha$-helix parameters: axial rise $= 1.5$~\AA/residue, radius $= 2.25$~\AA, turn angle $= 100^\circ$/residue. Linker lengths: 6--12 residues. Helix axis direction randomized within $\pm 65^\circ$ of the original inter-fragment vector; rotational phase sampled uniformly over $[0, 2\pi]$. Sampling density: 40 variants per source backbone and linker length. Clash filter: $\mathrm{C\alpha}$ distance $> 3$~\AA\ between non-bonded pairs. Compactness filter: $R_g < 2.1 \times$ expected $R_g$. Diversity selection: three-dimensional binning over radius of gyration, inter-fragment contact count, and centroid distance.

\paragraph{Domain decomposition and retrieval.} Merizo is used for domain segmentation. Domain-level and full-chain retrieval are performed using Foldseek against CATH S40. A backbone is counted as domain-retrievable if any decomposed domain achieves alnTM${\geq}\tau$; the all-domain rate is reported in Table~\ref{tab:all_domain}.

\section{Retrievability Under Different TM-Score Normalizations}
\label{app:retrievability}

Table~\ref{tab:app_drr_passed} gives the full four-condition cross of search granularity (domain vs.\ full-chain) with TM-score normalization (alnTM vs.\ qTM).

\begin{table*}[t]
\centering
\caption{Retrievability (\%) under four search$\times$normalization conditions, computed over AF3-passed structures. Dom: Merizo-decomposed domain PDBs as queries. FC: intact full-chain as query. alnTM: normalized by aligned region length. qTM: normalized by full query length. The Dom alnTM and FC qTM columns at threshold 0.5 correspond to the AF3-passed columns in Table~\ref{tab:app_af3_ablation}. Merizo domains average 122--138 residues across learned generators (111 for RetFold), close to the CATH S40 median of 132. The columns should be read as a two-by-two score diagnostic rather than as an additive decomposition.}
\setlength{\tabcolsep}{3mm}
\small
\begin{tabular}{lrrrrrrrr}
\toprule
 & \multicolumn{4}{c}{TM${\geq}0.5$} & \multicolumn{4}{c}{TM${\geq}0.7$} \\
\cmidrule(lr){2-5} \cmidrule(lr){6-9}
Method & Dom alnTM & Dom qTM & FC alnTM & FC qTM & Dom alnTM & Dom qTM & FC alnTM & FC qTM \\
\midrule
BoltzGen & 90.7 & 77.1 & 92.2 & 73.9 & 77.3 & 51.0 & 73.4 & 41.2 \\
PXDesign & 97.9 & 95.3 & 96.9 & 80.8 & 93.2 & 83.7 & 90.3 & 60.1 \\
RFDiffusion & 95.7 & 42.4 & 96.6 & 11.0 & 85.4 & 8.8 & 74.7 & 3.0 \\
ProtPardelle & 95.0 & 68.1 & 90.8 & 0.0 & 73.0 & 12.1 & 53.9 & 6.4 \\
FrameDiff & 78.3 & 49.3 & 94.2 & 1.4 & 50.7 & 14.5 & 42.0 & 13.0 \\
Chroma & 98.0 & 67.7 & 100.0 & 26.3 & 78.8 & 11.1 & 71.7 & 2.0 \\
FoldFlow & 82.9 & 35.8 & 43.4 & 0.0 & 43.6 & 4.9 & 16.9 & 0.2 \\
FrameFlow & 92.1 & 30.6 & 84.7 & 0.9 & 37.4 & 2.1 & 36.2 & 1.5 \\
\midrule
RetFold & 95.7 & 87.1 & 100.0 & 20.6 & 95.5 & 83.3 & 100.0 & 0.0 \\
\bottomrule
\end{tabular}
\label{tab:app_drr_passed}
\end{table*}

\subsection{Reference-Normalized, Coverage-Controlled Retrieval}

The primary DRR uses Foldseek alnTM, which normalizes similarity by the aligned region and is intentionally sensitive to local structural reuse. We add a stricter robustness analysis to distinguish a substantially covered known domain from a short high-quality local match. For each Merizo domain $d$ and its CATH S40 hit $r$, we exactly recompute target-normalized TM-score, denoted $\mathrm{tTM}(d,r)$, with the CATH domain as the normalization target. We also define target coverage as $\mathrm{tcov}(d,r)=L_{\mathrm{covered}}/L_r$. A strict match satisfies
\begin{equation}
\mathrm{StrictHit}(d,r)=
\mathbb{1}\!\left[
\mathrm{tTM}(d,r)\geq 0.5
\ \land\
\mathrm{tcov}(d,r)\geq 0.7
\right].
\end{equation}
Strict any-domain retrieval requires at least one strict match in a backbone, whereas strict all-domain retrieval requires every predicted domain to satisfy the same criterion. This analysis complements rather than replaces DRR: DRR measures whether a generated backbone reuses an alignable known structural unit, while the strict criterion asks whether that unit substantially covers a complete CATH reference domain.

\begin{figure*}[t]
\centering
\includegraphics[width=\textwidth]{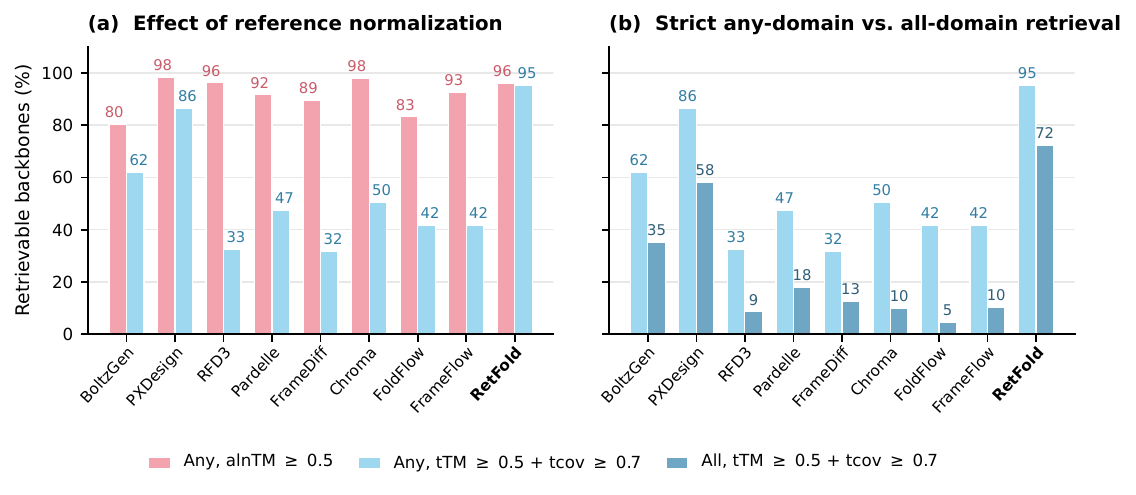}
\caption{Reference-normalized, coverage-controlled domain retrieval over the all-generated cohort using a fixed denominator of 500 backbones per method. (a)~Any-domain DRR under the primary aligned-length criterion (alnTM${\geq}0.5$) compared with strict any-domain retrieval, which requires tTM${\geq}0.5$ and tcov${\geq}0.7$. (b)~Strict any-domain compared with strict all-domain retrieval. The stricter criterion defines the boundary of the permissive DRR interpretation without replacing the primary domain--full-chain comparison.}
\label{fig:reference_normalized_drr}
\end{figure*}

Figure~\ref{fig:reference_normalized_drr} shows that strict any-domain retrieval is highest for RetFold (95.4\%) and PXDesign (86.4\%), followed by BoltzGen (62.0\%), Chroma (50.4\%), ProtPardelle (47.4\%), FoldFlow (41.8\%), FrameFlow (41.8\%), RFDiffusion (32.6\%), and FrameDiff (31.6\%). Strict all-domain retrieval is lower for several generators, including BoltzGen (35.2\%) and PXDesign (58.0\%). These results bound the main claim: aligned-length DRR identifies local reuse of database-matched structural units, whereas the strict criterion tests whether the matched reference domain is substantially covered.

We separately audited whether the $\mathrm{tcov}\geq0.7$ clause, rather than target-normalized TM-score, drives strict any-domain retrieval. Starting from $\mathrm{tTM}\geq0.5$, adding the coverage clause removes only 0.6--2.8 percentage points across the eight learned generators. Thus the large difference between aligned-length DRR and the strict analysis is driven primarily by the normalization convention; the additional coverage threshold is a comparatively small safeguard against incomplete reference-domain matches.

\subsection{Sensitivity to the Operational Protocol}

\begin{figure*}[t]
\centering
\includegraphics[width=\textwidth]{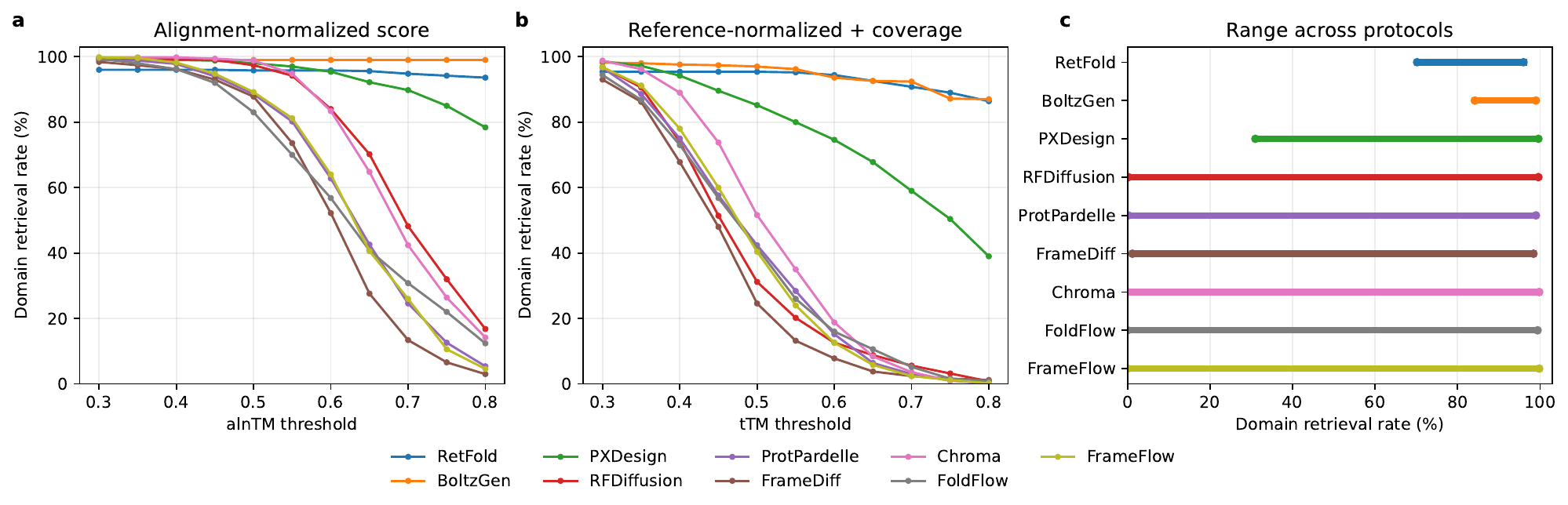}
\caption{Sensitivity of any-domain retrieval to the operational protocol over the all-generated cohort with a fixed denominator of 500 per method. (a)~DRR as the aligned-length TM-score threshold varies. (b)~DRR as the target-normalized TM-score threshold varies under target coverage $\geq0.7$. (c)~The range of DRR values across tested score, threshold, normalization, and coverage conventions. Wide ranges show that a binary novelty label is protocol dependent; they should not be interpreted as uncertainty in a single fixed estimator.}
\label{fig:operational_novelty_sensitivity}
\end{figure*}

We evaluated DRR over alnTM and target-normalized tTM thresholds from 0.30 to 0.80 in increments of 0.05. For tTM, we additionally varied target coverage from 0 to 0.9 and evaluated both reference-only and reciprocal coverage. Figure~\ref{fig:operational_novelty_sensitivity} shows that retrieval rates and model rankings change across these conventions. Aligned-length scores yield high rates over a broad threshold range, whereas reference-normalized scores with coverage control decline more sharply for several models. RetFold and BoltzGen remain high across much of the tested range, while other methods span substantially wider intervals.

This sensitivity does not invalidate structural retrieval. It defines what the measurement can support. The results establish protocol-relative statements of the form ``retrievable from database $\mathcal{R}$ under segmentation $S$, score $m$, threshold $\tau$, and coverage rule $c$.'' They do not provide a protocol-independent binary classification of whether a structure belongs to an exhaustively known fold space.

\subsection{Ablation: Effect of Designability Filtering}

Table~\ref{tab:app_af3_ablation} compares DRR computed over the all-generated cohort with DRR restricted to AF3-passed structures. The broad domain-retrieval pattern is retained, although the magnitude and direction of the shift are model dependent.

For the task-matched BoltzGen cohort, filtering increases DRR from 80.2\% to 90.7\% and FC from 62.0\% to 73.9\%; PXDesign changes more modestly, from 98.2\% to 97.9\% for DRR and from 75.4\% to 80.8\% for FC. Domain-level retrieval shifts by at most 1.5 points for six of the nine methods; the exceptions are BoltzGen ($-$10.5), FrameDiff ($+$11.1) and ProtPardelle ($-$3.4), all of which have small passed subsets. Full-chain retrieval rises under filtering for the low-pass-rate cohorts, most notably Chroma (7.4\% to 26.3\%, 99 passed). We therefore use the fixed all-generated cohort for the primary novelty analysis and treat the AF3-passed cohort as a complementary quality-conditioned view.

\begin{table*}[t]
\centering
\caption{Ablation under the primary top-100 protocol: DRR with vs.\ without designability filtering, where $\Delta = \text{All} - \text{AF3-passed}$. All-generated rates use a fixed denominator of 500 per method. AF3-passed rates use the passed subset shown in parentheses and are obtained by filtering the same domain- and full-chain-hit tables used for the all-generated rates.}
\label{tab:app_af3_ablation}
\small
\setlength{\tabcolsep}{5.6mm}
\begin{tabular}{lrrrrrrr}
\toprule
Method & \multicolumn{3}{c}{DRR (TM${\geq}0.5$)} & & \multicolumn{3}{c}{FC (qTM${\geq}0.5$)} \\
\cmidrule(lr){2-4} \cmidrule(lr){6-8}
 & All ($N$) & AF3-passed & $\Delta$ & & All ($N$) & AF3-passed & $\Delta$ \\
\midrule
BoltzGen & 80.2 (500) & 90.7 (410) & $-$10.5 & & 62.0 & 73.9 & $-$11.9 \\
PXDesign & 98.2 (500) & 97.9 (381) & +0.3 & & 75.4 & 80.8 & $-$5.4 \\
RFDiffusion & 96.2 (500) & 95.7 (328) & +0.5 & & 9.6 & 11.0 & $-$1.4 \\
ProtPardelle & 91.6 (500) & 95.0 (141) & $-$3.4 & & 0.0 & 0.0 & 0.0 \\
FrameDiff & 89.4 (500) & 78.3 (69) & +11.1 & & 0.4 & 1.4 & $-$1.0 \\
Chroma & 98.0 (500) & 98.0 (99) & 0.0 & & 7.4 & 26.3 & $-$18.9 \\
FoldFlow & 83.2 (500) & 82.9 (491) & +0.3 & & 0.0 & 0.0 & 0.0 \\
FrameFlow & 92.6 (500) & 92.1 (340) & +0.5 & & 0.6 & 0.9 & $-$0.3 \\
RetFold & 96.0 (500) & 95.7 (466) & +0.3 & & 20.0 & 20.6 & $-$0.6 \\
\bottomrule
\end{tabular}
\end{table*}

\subsection{Per-Length Results for All Models}

\begin{figure*}[t]
\centering
\includegraphics[width=\textwidth]{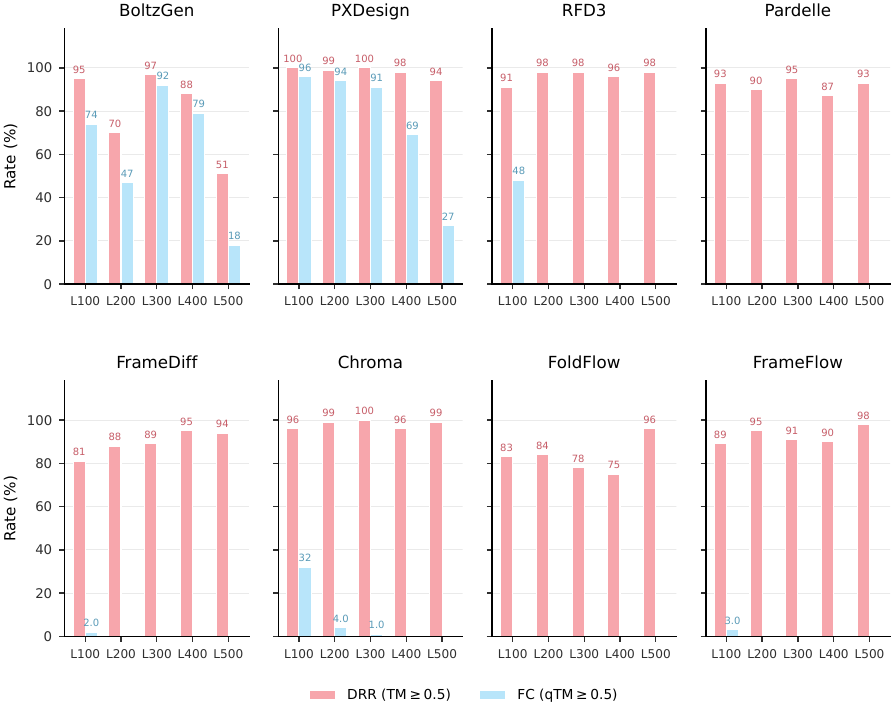}
\caption{Per-length DRR vs.\ FC for all eight learned generators under the primary top-100 protocol (all generated structures, no AF3 filtering). Red bars: DRR at alnTM${\geq}0.5$; blue bars: FC at qTM${\geq}0.5$. BoltzGen and PXDesign use the task-matched target-free generation settings described in the experimental setup. Length dependence is model specific; RFD3 denotes RFDiffusion and Pardelle denotes ProtPardelle.}
\label{fig:app_per_length_all}
\end{figure*}

Figure~\ref{fig:app_per_length_all} extends the all-generated analysis to each target length for all eight learned generators. DRR remains above 50\% in every model--length cell, but the full-chain trend is model dependent. Under the task-matched target-free runs, PXDesign FC decreases from 96\% at L100 to 27\% at L500, while BoltzGen decreases overall from 74\% to 18\% but is non-monotonic at intermediate lengths. RFDiffusion, Chroma, FoldFlow, and FrameFlow have low or rapidly decreasing FC at longer lengths. Thus, the domain--full-chain separation widens with length for several models rather than constituting a universal monotonic law. Each cell uses a fixed planned denominator of 100 generated structures.

\subsection{Domain Count Does Not Explain the Length Trend}

Because longer backbones are also segmented into more domains, we tested whether the full-chain trend could be attributed specifically to domain count. We fitted a logistic model for FC qTM${\geq}0.5$ over all eight learned generators, with Merizo domain count as the covariate of interest and model identity and the five target-length conditions as categorical fixed effects. Backbones for which Merizo returned no domain were excluded (11/4000). Uncertainty was estimated with 1000 bootstrap resamples within model--length strata. After adjustment, an additional predicted domain did not reduce full-chain retrieval (odds ratio 1.23, 95\% bootstrap interval 1.03--1.54). The pooled decline across unadjusted domain-count bins is therefore confounded by model and length composition. This is consistent with the query-length ceiling of Eq.~\ref{eq:ceiling}: target length, not the number of predicted units, sets the attainable maximum. We use the per-length analysis descriptively and do not claim that observed Merizo domain count is an independent causal mechanism.

\section{RetFold Length-Wise Behavior and Cost}
\label{app:length-cost}

\begin{figure*}[t]
\centering
\includegraphics[width=\textwidth]{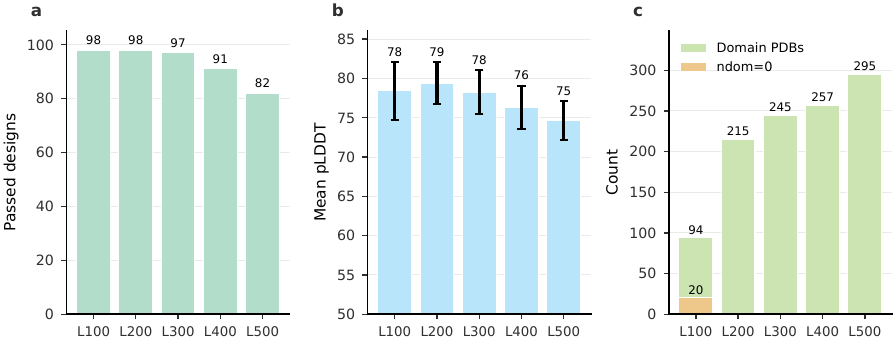}
\caption{RetFold length-wise quality-screen pass rate and domain-level retrieval. Any-domain retrieval at alnTM${\geq}0.7$ is lower at L100 (78.6\%), whereas the quality-screen pass count is lowest at L500.}
\label{fig:app_retfold_length}
\end{figure*}

Table~\ref{tab:app_retfold_length} and Figure~\ref{fig:app_retfold_length} report the length-wise quality-screen and retrieval profiles of the selected RetFold outputs. Domain retrieval is lower at L100 under the stricter threshold, while the all-domain rate also decreases modestly at L300 and L500.

\begin{table*}[t]
\centering
\caption{RetFold length-wise quality-screen counts and domain retrieval. Any-domain retrieval counts a selected output as retrievable if at least one Merizo domain matches CATH S40 at TM${\geq}0.7$; all-domain requires every Merizo domain to match at the same threshold. Retrieval percentages use the analyzable, quality-screened subset at each length.}
\small
\setlength{\tabcolsep}{10mm}
\begin{tabular}{lrrrr}
\toprule
Length & Generated & AF3 passed & Any-domain 0.7 & All-domain 0.7 \\
\midrule
L100 & 100 & 98 & 78.6 & 78.6 \\
L200 & 100 & 98 & 100.0 & 100.0 \\
L300 & 100 & 97 & 100.0 & 95.9 \\
L400 & 100 & 91 & 100.0 & 100.0 \\
L500 & 100 & 82 & 100.0 & 97.6 \\
\bottomrule
\end{tabular}
\label{tab:app_retfold_length}
\end{table*}

\subsection{Junction-Resolved RetFold Confidence}

\begin{figure*}[t]
\centering
\includegraphics[width=\textwidth]{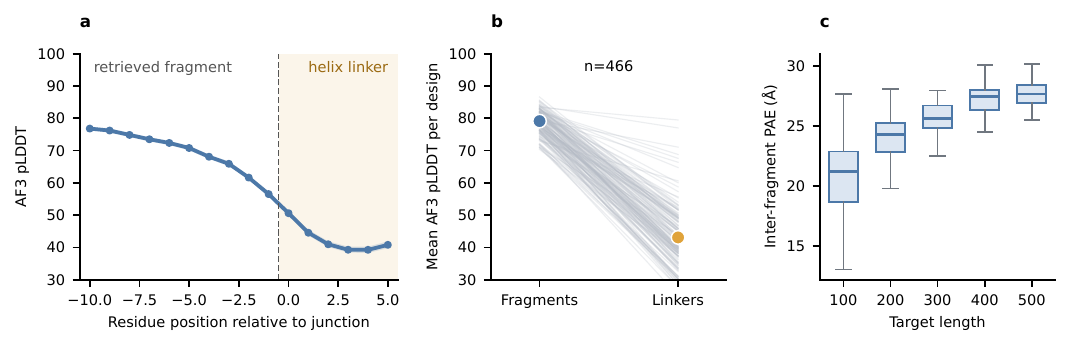}
\caption{Junction-resolved AF3 confidence for 466 RetFold backbones passing the whole-chain pLDDT threshold. (a)~Mean per-residue pLDDT aligned at each fragment--linker boundary; shading denotes a 95\% bootstrap confidence interval over designs. Negative positions lie in retrieved fragments and non-negative positions in inserted helix linkers. (b)~Within-design comparison of mean pLDDT for retrieved-fragment and linker residues; gray lines show a reproducible random subset of 140 designs and colored points show means over all 466. (c)~Inter-fragment PAE by target length; boxes show the median and interquartile range, with whiskers extending to 1.5 times the interquartile range. Low linker confidence and high inter-fragment PAE identify the junctions as RetFold's principal unresolved designability limitation.}
\label{fig:app_junction_confidence}
\end{figure*}

To test the part of RetFold that is not guaranteed by retrieval, we mapped the exact Stage~2 helix-linker positions from the construction manifest onto the best-of-eight AF3 prediction for each of the 466 backbones that passed the whole-chain pLDDT threshold. Per-residue pLDDT was averaged over atoms within each residue. We compared the two inserted linkers with the retrieved-fragment residues within each design and computed PAE between each pair of retrieved fragments. Confidence decreases as the sequence approaches a fragment--linker boundary and remains low throughout the inserted linker (Figure~\ref{fig:app_junction_confidence}a). Across designs, mean linker pLDDT is 43.1 versus 79.1 in retrieved fragments, a paired mean difference of $-36.0$ (95\% design-bootstrap CI $[-36.8,-35.2]$; two-sided Wilcoxon signed-rank $P<10^{-77}$). Mean inter-fragment PAE is 24.9\,\AA\ and increases with target length. These diagnostics show that passing a whole-chain average-pLDDT threshold does not validate the assembled junctions or the relative placement of retrieved domains.

Table~\ref{tab:app_timing} reports the generation-only cost underlying the speedup quoted in the main text. The two RetFold rows separate Stage~1 retrieval and assembly from the complete Stage~1$+$2 pipeline; all reported RetFold results, and the speedup comparison, use the latter. Seconds per backbone is the observed elapsed time divided by the number of generated structures, and wall time is this mean multiplied by 500. For a consistent device-level description, hardware denotes the device assigned to one generation worker; independent workers could run concurrently. The figures are therefore per-backbone wall-time averages rather than GPU-second or hardware-normalized measurements.

\begin{table*}[t]
\centering
\caption{Generation-only timing, ordered by observed mean cost per backbone. ProteinMPNN and AlphaFold3 validation are excluded because they are shared downstream steps. Hardware is reported per generation worker; total concurrent worker counts differ across runs. Seconds per backbone is an elapsed-time mean, not a GPU-second or hardware-normalized measurement.}
\small
\setlength{\tabcolsep}{6.2mm}
\begin{tabular}{lrrrl}
\toprule
Method & Generated & Wall time (h) & Seconds/backbone & Hardware per worker \\
\midrule
RetFold (Stage 1 only) & 500 & 0.007 & 0.05 & Intel Xeon Gold 6530 CPU \\
RetFold (Stage 1$+$2) & 500 & 0.275 & 1.98 & Intel Xeon Gold 6530 CPU \\
ProtPardelle & 500 & 0.037 & 1.06 & 1$\times$NVIDIA H200 \\
BoltzGen & 500 & 0.257 & 5.52 & 1$\times$NVIDIA H200 \\
PXDesign & 500 & 1.060 & 7.63 & 1$\times$NVIDIA H200 \\
FoldFlow & 500 & 1.175 & 8.46 & 1$\times$NVIDIA H200 \\
Chroma & 500 & 3.600 & 25.95 & 1$\times$NVIDIA H200 \\
FrameDiff & 500 & 6.430 & 46.52 & 1$\times$NVIDIA H200 \\
FrameFlow & 500 & 19.100 & 137.53 & 1$\times$NVIDIA H200 \\
RFDiffusion & 500 & 30.370 & 218.67 & 1$\times$NVIDIA H200 \\
\bottomrule
\end{tabular}
\label{tab:app_timing}
\end{table*}

\section{Diversity and Coverage}
\label{app:diversity}

Table~\ref{tab:app_diversity_coverage} gives the underlying values for the diversity and coverage comparison in the main text. The two quantities measure different things: mean pairwise qTM is computed among a method's own designable outputs, so a low value means those outputs are unlike one another, while the CATH counts measure how much of the known classification those outputs touch.

Both quantities depend on how many structures are available, and that number varies substantially across methods because it is set by each method's AF3 pass rate rather than by design. Coverage is the more affected quantity because topology and superfamily counts are cumulative. We therefore rarefied each current AF3-passed hit set to the common minimum of 69 structures, repeated sampling 1000 times, and counted the union of CATH topologies retrieved at domain alnTM${\geq}0.7$. RetFold covers a mean of 36.8 topologies (95\% bootstrap interval 31--42), while RFDiffusion covers 36.5 (29--44); their intervals overlap substantially. BoltzGen, PXDesign, ProtPardelle, FrameDiff, Chroma, FoldFlow, and FrameFlow yield means of 23.2, 25.9, 26.6, 14.0, 26.8, 14.7, and 14.3, respectively. The raw counts below record demonstrated coverage rather than a sample-size-normalized ranking. Within-length all-versus-all Foldseek covers all ordered non-self pairs for the task-matched BoltzGen and PXDesign cohorts and yields mean pairwise qTM values of 0.435 and 0.270.

\begin{table*}[t]
\centering
\caption{Structural diversity and CATH coverage among analyzable designable outputs. Lower mean pairwise qTM indicates greater structural diversity. Pairwise qTM is computed among AF3-passed backbones within each target-length bin and pooled over all ordered non-self pairs. Structure counts differ across methods because they follow each method's AF3 pass rate; coverage counts are cumulative and therefore not directly comparable across rows. Coverage uses AF3-passed domain hits at alnTM${\geq}0.7$ and the CATH domain classification.}
\small
\setlength{\tabcolsep}{7.4mm}
\begin{tabular}{lrrrr}
\toprule
Method & Structures & Mean pairwise qTM & CATH topologies & CATH superfamilies \\
\midrule
BoltzGen & 410 & 0.435 & 63 & 104 \\
PXDesign & 381 & 0.270 & 62 & 97 \\
RFDiffusion & 328 & 0.206 & 136 & 236 \\
ProtPardelle & 141 & 0.174 & 62 & 93 \\
FrameDiff & 69 & 0.207 & 19 & 28 \\
Chroma & 99 & 0.213 & 42 & 67 \\
FoldFlow & 491 & 0.170 & 57 & 148 \\
FrameFlow & 340 & 0.159 & 56 & 103 \\
RetFold & 466 & 0.179 & 82 & 143 \\
\bottomrule
\end{tabular}
\label{tab:app_diversity_coverage}
\end{table*}

\section{Robustness to Domain Segmentation Method}
\label{app:segmentation}

To quantify sensitivity to the domain segmentation rule, we repeated the domain-level DRR analysis using Chainsaw~\cite{wells2024chainsaw}, a fully convolutional neural network for protein domain boundary prediction developed by the CATH team. Chainsaw is trained directly on CATH domain assignments and was used in the TED (The Encyclopaedia of Domains) project to parse AlphaFold Database structures. We applied Chainsaw to the same all-generated cohort used in the Merizo analysis, retained the fixed denominator of 500 backbones per method, and recomputed domain-level DRR at alnTM${\geq}0.5$ and alnTM${\geq}0.7$ under the same CATH S40 Foldseek protocol.

Table~\ref{tab:app_segmentation_robustness} compares domain-level DRR obtained with Merizo and Chainsaw. Chainsaw assigns fewer domains per backbone for every learned generator and lowers DRR@0.5 for seven of the eight learned methods. The change is small for BoltzGen and PXDesign ($+2.4$ and $-3.0$~pp), but reaches $-44.0$~pp for FoldFlow. Under Chainsaw, DRR@0.5 ranges from 39.2\% for FoldFlow to 99.8\% for RetFold; eight of nine methods remain above 50\%. At alnTM${\geq}0.7$, the largest change is $-30.8$~pp for Chroma. RetFold provides the one case where the correct segmentation is known: its backbones contain three fragments by construction, and both tools report fewer (2.43 for Merizo, 2.27 for Chainsaw), so segmentation loss is present even where the units are unmodified reference domains. These results show that database-matched structure remains detectable under an independent segmentation, but the estimated prevalence and model ordering are segmentation-dependent. They therefore support protocol-relative reporting rather than a segmentation-invariant novelty classification.

\begin{table*}[t]
\centering
\caption{Domain-level DRR comparison between Merizo and Chainsaw on the all-generated cohort ($N=500$ per method). Domain counts use the same fixed denominator; the two absent FrameDiff structures contribute zero domains and fail retrieval. $\Delta=\mathrm{Chainsaw}-\mathrm{Merizo}$ in percentage points. Both segmentations use the same CATH S40 search and alnTM thresholds.}
\setlength{\tabcolsep}{4mm}
\begin{tabular}{lrrrrrrrr}
\toprule
Method & \multicolumn{2}{c}{Domains/backbone} & \multicolumn{3}{c}{DRR (alnTM${\geq}0.5$)} & \multicolumn{3}{c}{DRR (alnTM${\geq}0.7$)} \\
\cmidrule(lr){2-3} \cmidrule(lr){4-6} \cmidrule(lr){7-9}
 & Merizo & Chainsaw & Merizo & Chainsaw & $\Delta$ & Merizo & Chainsaw & $\Delta$ \\
\midrule
BoltzGen     & 2.00 & 1.00 & 80.2 & 82.6 & +2.4 & 67.8 & 62.8 & $-$5.0 \\
PXDesign     & 2.14 & 1.34 & 98.2 & 95.2 & $-$3.0 & 94.4 & 90.8 & $-$3.6 \\
RFDiffusion  & 2.30 & 1.43 & 96.2 & 92.4 & $-$3.8 & 65.2 & 42.2 & $-$23.0 \\
ProtPardelle & 2.38 & 1.57 & 91.6 & 76.4 & $-$15.2 & 49.0 & 27.0 & $-$22.0 \\
FrameDiff    & 2.15 & 1.45 & 89.4 & 80.0 & $-$9.4 & 30.2 & 16.8 & $-$13.4 \\
Chroma       & 2.38 & 1.65 & 98.0 & 83.0 & $-$15.0 & 58.2 & 27.4 & $-$30.8 \\
FoldFlow     & 2.46 & 1.43 & 83.2 & 39.2 & $-$44.0 & 37.2 & 14.2 & $-$23.0 \\
FrameFlow    & 2.33 & 1.52 & 92.6 & 73.4 & $-$19.2 & 43.0 & 25.0 & $-$18.0 \\
RetFold      & 2.43 & 2.27 & 96.0 & 99.8 & +3.8 & 95.8 & 99.8 & +4.0 \\
\bottomrule
\end{tabular}
\label{tab:app_segmentation_robustness}
\end{table*}

\section{Criterion Calibration}
\label{app:calibration}

The retrieval rates reported in the main text are conditional on a similarity
score and a threshold. This appendix calibrates the three scores we use in both
directions: sensitivity, using a positive control whose composition is known by
construction, and specificity, using a held-out-topology negative control.

\subsection{Query-Length Ceiling}

Eq.~\ref{eq:ceiling} in the main text states the bound.
CATH S40 contains domains rather than chains (median length 132 residues,
interquartile range 92--194), so a generated chain must find a reference domain
at least half its own length before $\mathrm{qTM}\geq0.5$ is attainable at all.
Table~\ref{tab:app_eligibility} reports the fraction of CATH S40 entries that
satisfy this necessary condition at each target length. The eligible fraction
falls from 95.95\% at $L{=}100$ to 13.39\% at $L{=}500$. This is a property of
the reference database and the score definition alone; it involves no generated
structure.

\begin{table}[ht]
\centering
\caption{Fraction of CATH S40 entries long enough to permit $\mathrm{qTM}\geq0.5$
against a query of the stated length, i.e.\ $L_r \geq 0.5\,L_q$.}
\small
\setlength{\tabcolsep}{4mm}
\begin{tabular}{lr}
\toprule
Query length & Eligible entries (\%) \\
\midrule
100 & 95.95 \\
200 & 70.10 \\
300 & 41.02 \\
400 & 23.60 \\
500 & 13.39 \\
\bottomrule
\end{tabular}
\label{tab:app_eligibility}
\end{table}

The ceiling is necessary but not sufficient: an eligible reference must also
align well. Empirically the decline is steeper than
Table~\ref{tab:app_eligibility} alone predicts. For RFDiffusion, full-chain
$\mathrm{qTM}\geq0.5$ retrieval falls from 56\% at $L{=}100$ to 1\% at $L{=}200$
and to approximately 0\% beyond, whereas the target-normalized
$\mathrm{tTM}\geq0.5$ rate over the same structures is 61\%, 52\%, 33\%, 25\%
and 32\%. Since $\mathrm{tTM}$ normalizes by the reference length, it carries no
query-length ceiling; the residual decline under $\mathrm{tTM}$ is the part of
the length trend that is not attributable to the normalization convention.

We note that $\mathrm{alnTM}$ and $\mathrm{qTM}$ are not the same alignment under
two normalizations. Each is optimized separately, and individual hits with
$\mathrm{qTM}>\mathrm{alnTM}$ occur. Differences between the two columns of
Table~\ref{tab:app_drr_passed} therefore reflect both a change of denominator and
a change of the alignment that maximizes the score.

\subsection{Positive Control}

RetFold backbones are assembled from unmodified CATH S40 domains, so the correct
answer for any retrieval criterion applied to them is known by construction: each
chain contains at least one complete reference domain. Searching the generated
backbones (not their AlphaFold3 re-predictions) against CATH S40 under
target-normalized TM-score recovers this: all 500 backbones exceed
$\mathrm{tTM}\geq0.5$, $\geq0.7$ and $\geq0.9$, with best-hit
$\mathrm{tTM}$ mean 0.999, median 1.000, fifth percentile 0.992 and minimum
0.967. The criterion is therefore not marginally satisfied on this control but
saturated.

Under the conventional full-chain $\mathrm{qTM}\geq0.5$ protocol the same 500
backbones are retrieved at 20.0\%. The 80-point difference is a property of the
normalization, since the reference domains are present in every chain by
construction.

\subsection{Negative Control}

To estimate specificity we require queries whose correct answer is
\emph{not} retrievable. We performed an all-versus-all search of CATH S40
against itself and then, for each query domain, discarded every hit belonging to
the query's own CATH topology before taking the best remaining hit. For such a
query the fold class is absent from the searched set, so any hit that passes
threshold is a false positive. We use the topology level because
$\mathrm{TM}>0.5$ is conventionally read as a same-fold
criterion~\cite{xu2010significant}; excluding at
the homologous-superfamily level would count remote homologues of the same fold
as errors.

Three points follow. First, $\mathrm{alnTM}$ at $\tau{=}0.5$ retrieves 90.04\%
of queries whose fold class has been removed from the searched set; at this
threshold it carries little discriminative information, and its high rates in
the main text should be read accordingly. Second, $\mathrm{qTM}$ attains the
highest AUC, but this calibration uses domain-sized queries against
domain-sized references. Under Eq.~\ref{eq:ceiling} its behaviour on multi-domain
chains is not represented here, and the positive control above shows its
sensitivity there is 20\%. High AUC on matched-length queries and low
sensitivity on long queries are consistent. Third, $\mathrm{tTM}$ is the only
score in this set that is saturated on the positive control while reaching a
false positive rate below 10\%, and it does so only at $\tau{=}0.7$ (5.38\%,
with no loss on the positive control); at $\tau{=}0.5$ its rate of 29.97\% is
close to the domain-level $\mathrm{qTM}$ rate and offers little margin.

False positive rates vary with structural class. At $\mathrm{tTM}\geq0.5$ they
are 36.2\% for mainly-$\alpha$, 24.7\% for mainly-$\beta$, 30.0\% for
$\alpha/\beta$, 13.6\% for few-secondary-structure and 20.4\% for special
classes, consistent with the lower topological degeneracy of irregular
structures. We do not convert this into model-specific corrections, because a
generated backbone has no ground-truth CATH class and using the class of its
best hit is circular when the hit itself may be the false positive.

\subsection{Reference-Normalized Retrieval Across Methods}

Table~\ref{tab:app_ttm_threshold} in the main text applies the calibrated
criterion to full-chain queries. We report both thresholds and do not interpret
the unretrieved remainder as a count of new folds. The false-positive rates in
Table~\ref{tab:app_loto} are measured on domain-sized queries and bound that
query population; we therefore do not use them to judge whether a given
chain-sized retrieval rate exceeds chance. The criterion cannot resolve whether
the unretrieved fraction contains new folds, distorted known folds, or both.

\subsection{Noise Floor Details}

The reference value of $\Delta$ and the observed per-method values are given in the main text.
Two caveats bound that comparison. The false positive rates are estimated on
domain-sized queries, which match the query population of DRR but not that of FC,
where Eq.~\ref{eq:ceiling} suppresses $\mathrm{qTM}$ further and the true floor is
correspondingly higher. And the leave-one-topology-out exclusion removes a
discrete label rather than a region of structure space, so cross-topology hits at
$\mathrm{TM}>0.5$ include genuine structural similarity; the reported rates are
therefore upper bounds on the false positive rate rather than point estimates.

\end{document}